\documentclass[aps,pre,twocolumn,groupedaddress,superscriptaddress,showpacs]{revtex4-1}

\usepackage{graphicx}
\usepackage{color}
\usepackage{amsmath}
\usepackage{amsfonts}
\usepackage{amssymb}
\usepackage{dcolumn}
\usepackage{float}
\usepackage{subfigure}

\definecolor{MyDarkBlue}{rgb}{0,0.08,0.45}
\definecolor{MyDarkGreen}{rgb}{0,0.45,0.08}
\definecolor{MyDarkRed}{rgb}{0.45,0,0.08}

\begin{document}
  \title{Optimal path cracks in correlated and uncorrelated lattices}

  \author{E. A. Oliveira}
    \email{erneson@fisica.ufc.br}
    \affiliation{Departamento de F\'isica, Universidade Federal do Cear\'a, Campus do Pici, 60451-970 Fortaleza, Cear\'a, Brazil}

  \author{K. J. Schrenk}
    \email{jschrenk@ethz.ch}
    \affiliation{Computational Physics for Engineering Materials, IfB, ETH Zurich, Schafmattstr. 6, 8093 Zurich, Switzerland}

  \author{N. A. M. Ara\'ujo}
    \email{nuno@ethz.ch}
    \affiliation{Computational Physics for Engineering Materials, IfB, ETH Zurich, Schafmattstr. 6, 8093 Zurich, Switzerland}

  \author{H. J. Herrmann}
	\email{hans@ifb.baug.ethz.ch}
	\affiliation{Departamento de F\'isica, Universidade Federal do Cear\'a, Campus do Pici, 60451-970 Fortaleza, Cear\'a, Brazil}
	\affiliation{Computational Physics for Engineering Materials, IfB, ETH Zurich, Schafmattstr. 6, 8093 Zurich, Switzerland}

  \author{J. S. Andrade Jr.}
	\email{soares@fisica.ufc.br}
	\affiliation{Departamento de F\'isica, Universidade Federal do Cear\'a, Campus do Pici, 60451-970 Fortaleza, Cear\'a, Brazil}
	\affiliation{Computational Physics for Engineering Materials, IfB, ETH Zurich, Schafmattstr. 6, 8093 Zurich, Switzerland}


  \begin{abstract}
    The optimal path crack model on uncorrelated surfaces, recently introduced by Andrade {\it et al.} ({\it Phys. Rev. Lett.} \textbf{103}, 225503, 2009), is studied in detail and its main percolation exponents computed. In addition to $\beta/\nu = 0.46 \pm 0.03$ we report, for the first time, $\gamma/\nu = 1.3 \pm 0.2$ and $\tau = 2.3 \pm 0.2$.
    The analysis is extended to surfaces with spatial long-range power-law correlations, where non-universal fractal dimensions are obtained when the degree of correlation is varied.
    The model is also considered on a three-dimensional lattice, where the main crack is found to be a surface with a fractal dimension of $2.46\pm0.05$.
  \end{abstract}

  \maketitle

  \section{Introduction}

Finding the optimal path between two points in a disordered system is a relevant challenge for science and technology \cite{Mezard84,Ansari85,Huse85,Huse85a,Kirkpatrick85,Kardar86,Kardar87,Perlsman92,Kertesz93,Perlsman96,Havlin05}.
This optimization problem is present in our daily lives when, for example, we make use of the {\it Global Positioning System} (GPS) to trace the best route to arrive to our destination.
This problem, however, is not only relevant for human transportation.
In materials science, the characterization of the optimal path is of extreme importance to study fractures, polymers in random environments, and transport in porous media \cite{Schwartz98}.
If intensively used, this path is prone to fail, and a new path needs to be found.
Studying how the successive paths evolve until the final configuration -- where connectivity is no longer possible -- is a challenge in itself which we address in this work.

Recently, Andrade {\it et al.}\ \cite{Andrade09} introduced a new model, named optimal path crack (OPC), to study the evolution of successive optimal paths under constant failure.
They have shown that, if a disordered energy landscape is considered, and each optimal path fails at its maximum energy site, the cracking process leads to a configuration where no more paths can be found.
For uncorrelated systems, regardless the degree of disorder, the shortest path in the main crack -- the minimal one to interrupt the connectivity -- is always a self-similar object with a fractal dimension of $1.22\pm0.02$.
This fractal dimension has been reported in several different systems like, e.\,g., the watershed line \cite{Fehr09,*Fehr11} and the perimeter of the percolative cluster at a discontinuous transition \cite{Araujo10}.
In this work, we first study in detail the properties of the OPC in uncorrelated lattices.
The fractal dimensions are accurately obtained and, being a percolation-like process, the main critical exponents of percolation are computed for the crack at the final adsorbing state \cite{Stauffer94,Sahimi94}. Besides the previously reported value of $\beta/\nu$, in this work we show that $\gamma/\nu = 1.3 \pm 0.2$ and the Fisher exponent is $\tau = 2.3 \pm 0.2$.
We then calculate and discuss the main properties of the model in three dimensions.

The interplay between correlation and randomness lies at the very core of emergent phenomena.
Here we also extend the study of the OPC to correlated lattices.
As in many previous studies \cite{Prakash92,Sahimi94a,Sahimi96,Makse96,Kikkinides99,Stanley99,Makse00,Araujo02,Araujo03,Du04}, spatial long-range correlated energy distributions on the lattice are accessed by fractional Brownian motion (FBM) \cite{Peitgen88} -- a generalized version of the classical Brownian motion, introduced by Mandelbrot and Ness \cite{Mandelbrot68} -- where the degree of correlation between the successive steps can be tuned.

The manuscript is organized as follows. 
In Section \ref{sec::model_and_definitions} the optimal path crack model is revised and its extension to correlated lattices is introduced.
The results for uncorrelated lattices, in the weak and strong disorder regime, are presented and discussed in Sec.~\ref{sec::uncorrelated_lattices} together with the main results for the three-dimensional system.
The behavior of the optimal path crack in correlated lattices is analyzed in Sec.~\ref{sec::correlated_lattices}.
We leave the final remarks for Sec.~\ref{sec::summary}.

  \section{\label{sec::model_and_definitions}Model and Definitions}
\subsection{\label{sec::op_gen}The Optimal Path} 
 \begin{figure}
   \includegraphics[width=\columnwidth]{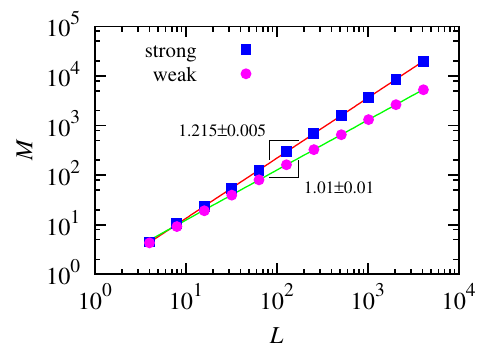}
   \caption{
     (Color online) Double logarithmic plot of the mass of the optimal path, $M_\text{op}$, in 2D as a function of the linear system size, $L$, for two different types of energy distribution. 
     The lower points ($\bullet$) are measured for uniformly random site energies, being equivalent to weak disorder. 
     The corresponding exponent is $d_\text{op}=1.01\pm 0.01$ and the optimal path is a self-affine object \cite{Porto99}.
     The upper line ($\blacksquare$) shows results obtained for site energies distributed according to a power-law, Eq.\,(\ref{eqn::power_dens}), with $\beta_D=1600$, such that the systems are in the strong disorder limit (for the considered system sizes). 
    In the strong disorder limit the optimal path is a self-similar object with a fractal dimension $d_\text{op}=1.215\pm 0.005$. 
    The optimal path search was performed using the Dijkstra algorithm \cite{Dijkstra59}. 
    Results have been averaged over $10^{10}$ samples for the smallest system size and $10^{4}$ for the largest one. 
    The error bars of the individual points are smaller than the symbols.
     \label{fig::first_cross}
   }
 \end{figure}
The optimal path crack (OPC) is obtained by systematically interrupting the optimal path (OP) on a random landscape.

For simplicity, let us consider a simple square lattice of linear size $L$ where an energy $\varepsilon_i \geq 0$ is assigned to every site. 
The energy of a path in a system is the sum over the energy of all sites in the path. 
The optimal path is the one, among all paths connecting two sites -- e.\,g., the bottom and the top of the lattice -- with the lowest energy. 
In the limit where all sites have the same energy, the OP is a straight line with mass (number of sites) $M_\text{op}=L$.

When the energies are randomly distributed, the OP is, in general, not a straight line -- its mass is larger or equal to $L$ -- and its properties depend on the energy distribution \cite{Cieplak94,Cieplak96,Porto97,Porto99}. 
In this work we take two energy distributions: uniform and infinite disorder (power law with exponent $-1$).
In the former case, the mass of the OP, $M_\text{op}$, scales linearly with the system size as seen in Fig.\,\ref{fig::first_cross}, while this is not the case in the infinite disorder limit.

When the special case of a power-law distribution with exponent $-1$ is considered,

\begin{equation}\label{eqn::power_dens}
	p(\varepsilon_i) \propto \frac{1}{\varepsilon_i} \ \ ,
\end{equation}

\noindent truncated between $\varepsilon_\text{max} = 1$ and $\varepsilon_\text{min} = \exp(-\beta_D)$, the disorder in the system is controlled by the disorder parameter $\beta_D\geq 0$. 
In the limit $\beta_D\to 0$, $\varepsilon_\text{min}\to\varepsilon_\text{max}$, and the OP approaches a straight line. 
For large values of $\beta_D$, the truncated distribution becomes very broad and the disorder strong (ultrametric limit).
The total energy of a given path is then mainly driven by the site with the highest energy in the path (see \cite{Cieplak94} and references therein). 
For a certain $\beta_D$, the system can be either in a strong or weak disorder regime, depending on its size. 
The crossover from weak to strong disorder has been investigated in detail by Porto \emph{et al.}\ \cite{Porto99}. 
For strong disorder, the OP is a self-similar object with a fractal dimension $d_\text{op}=1.215\pm 0.005$ (see Fig.\,\ref{fig::first_cross}).

The optimal path is typically the one heavily used in a system being liable to fail. 
This failure is most likely to occur in the highest energy site. 
Under failure, this site is destroyed and can no longer be used. 
To optimize the transportation ``cost'' on the modified substrate, a new OP has to be found, which is again disrupted at its highest energy site. 
During this process, transport in the system remains possible, with increasing cost, until the formation of a path of destroyed sites which disconnects the system into two parts. 
This process, denoted as \emph{optimal path cracking}, was introduced recently by Andrade \emph{et al.} \cite{Andrade09}, who discovered that the shortest path of destroyed sites necessary to disconnect the system has a fractal dimension of $1.22\pm 0.02$, an interesting exponent also found in several other systems \cite{Cieplak94,Cieplak96,Porto99,Fehr09,*Fehr11,Araujo10,Andrade11}.

    \subsection{\label{sec::opc_uncorr}Optimal Path Cracking}

The starting configuration for each realization of OPC is a regular square or cubic lattice of linear size $L$ where all sites and bonds are occupied (open). 
Notwithstanding the site definition of the problem, the model can be straightforwardly extended to consider bonds instead of sites. 
Periodic boundary conditions are applied in all directions, except for one, where fixed boundary conditions are taken. 
For definiteness, this direction is called vertical, such that there are no bonds directly connecting the lowest row (layer) and the top row (layer) in two (three) dimensions. 
For each realization, a non-negative energy value $\varepsilon_i$, $i=1,\dots,N=L^d$ (where $d$ is the dimension of the system), is assigned to each site, either distributed uniformly (e.\,g.\ in $[0,1)$) or according to a truncated power law (Eq.\,(\ref{eqn::power_dens})). 
In the latter case, applying the transformation method for distributions yields

\begin{equation}\label{eqn::energy_uncorr}
	\varepsilon_i = \exp\left[\beta_D (x_i-1)\right] \ \ ,
\end{equation}

\noindent where $x_i$ is a random number uniformly distributed in $[0,1)$.

Starting from this setup, the OPC is generated by the following procedure:

\begin{enumerate}
	\item  Find the OP through the lattice, connecting the lowest row and the highest one (the lowest layer and the top layer in 3D). Only occupied sites can be part of a path and only nearest neighbors are considered to be connected.
	\item Identify and remove the most vulnerable site in the optimal path, i.\,e., the one with the highest energy.
	\item Repeat steps 1 and 2 until there is no path connecting the bottom row (layer) to the top row (layer).
	\item The configuration of unoccupied sites obtained in this way is the OPC. 
\end{enumerate}

It is convenient to use Hoshen--Kopelmann-like labeling \cite{Hoshen76,Newman00,Newman01} to keep track of the properties of the clusters of removed sites, as the OPC emerges. 
The properties of the cracks are analyzed by monitoring three different masses:

\begin{itemize}
	\item The number of all removed sites in the system, $M_\text{tot}$ -- the density $\rho$ is then defined as $\rho=M_\text{tot} / N$;
	\item The number of blocked sites forming the largest cluster disconnecting the system, which is called $M_\text{lc}$;
	\item The shortest path in the largest cluster, sufficient to disconnect the system, with a mass $M_\text{sp}$.
\end{itemize}

\begin{figure*}
	\centering
	\subfigure[\label{fig::sample_0_002}]{		\includegraphics[width=1\columnwidth]{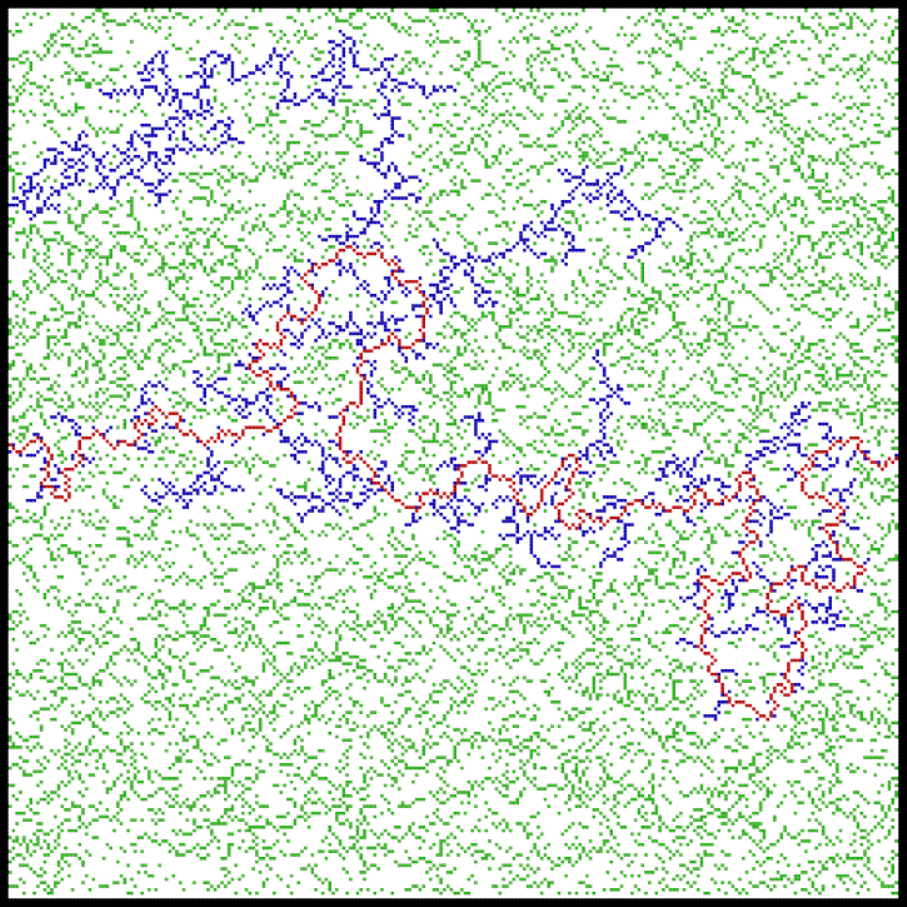}
	}
	\subfigure[\label{fig::sample_rand}]{		\includegraphics[width=1\columnwidth]{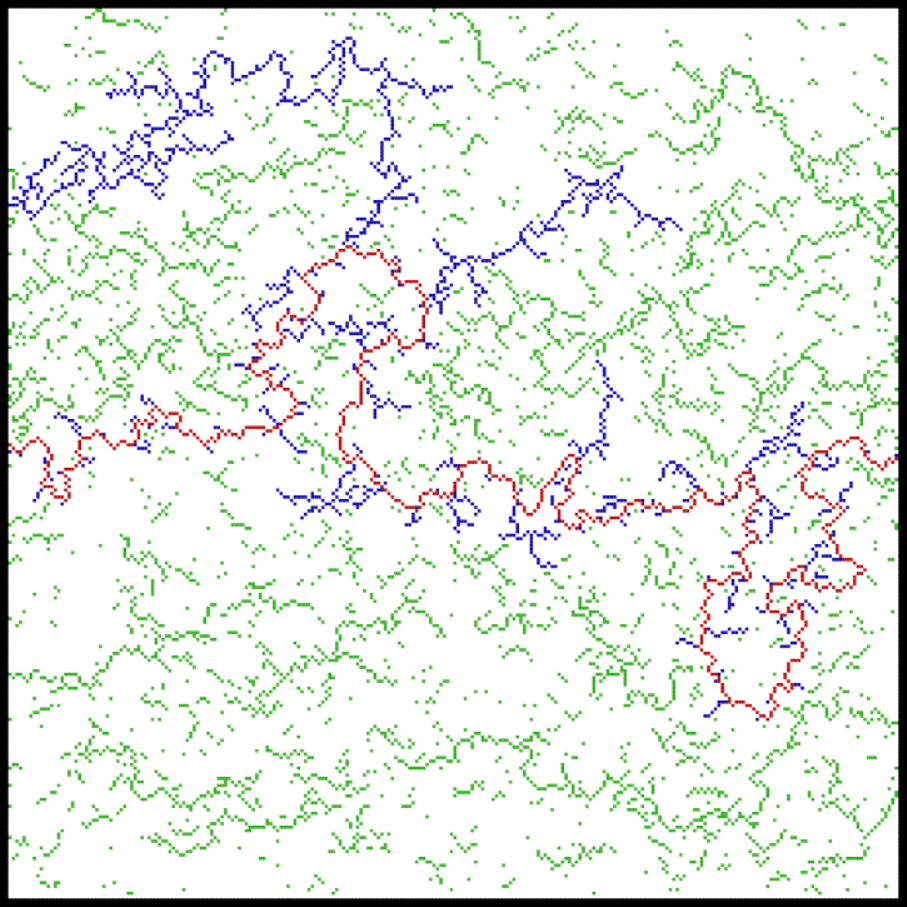}
	}
	\subfigure[\label{fig::sample_6}]{		\includegraphics[width=1\columnwidth]{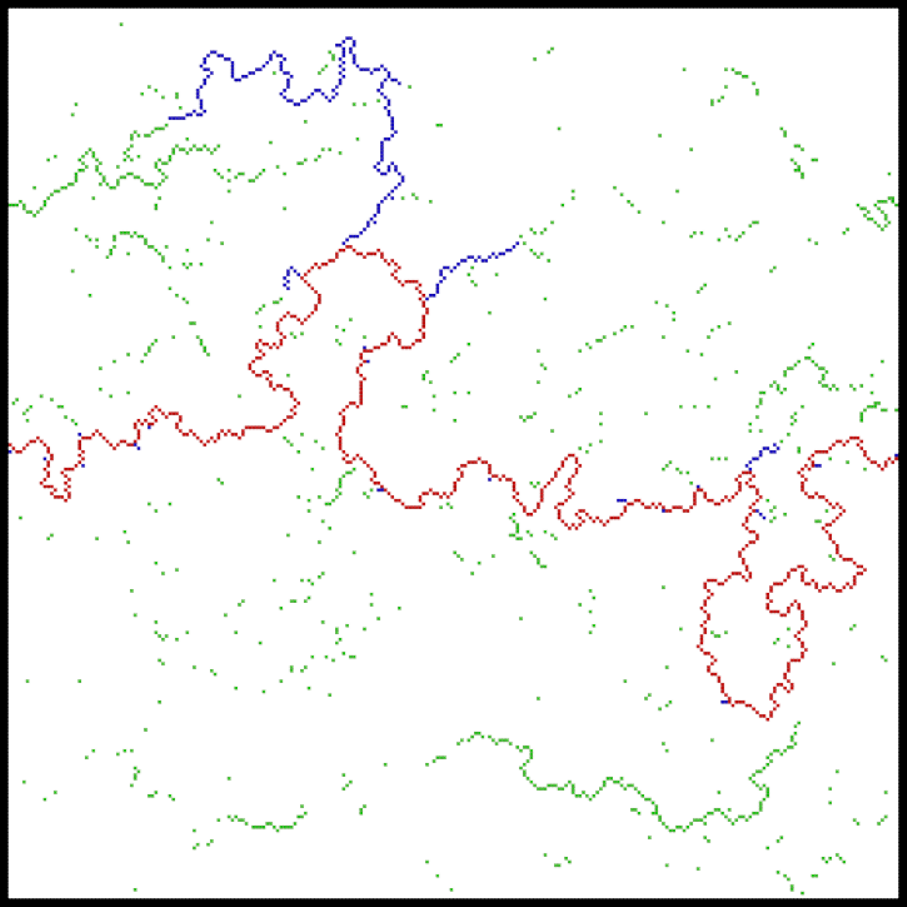}
	}
	\subfigure[\label{fig::sample_100}]{		\includegraphics[width=1\columnwidth]{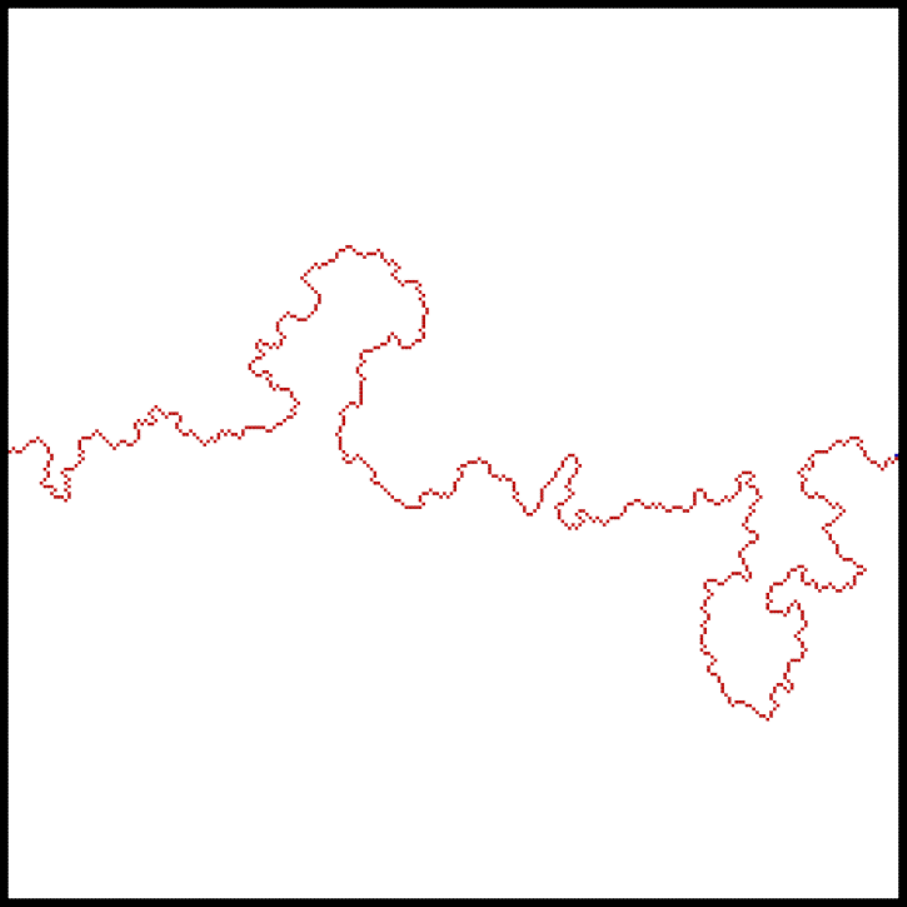}
	}
	\caption{\label{fig::cluster_samples}
	(Color online) Snapshots of optimal path crack clusters on uncorrelated lattices, generated with the same seed for the uniformly distributed pseudo-random numbers. 
        Figures (a) and (b) show equivalent realizations of the weak disorder limit: in (a) the site energies are distributed according to a power-law, Eq.\,(\ref{eqn::power_dens}), with $\beta_D=0.002$, while in (b) they are uniformly distributed.
       Figures (c) and (d) show snapshots for $\beta_D=6$ and $\beta_D=100$, respectively, corresponding to the intermediate and the strong disorder regime.
       The shortest path in the largest optimal path crack cluster is shown in red (medium gray), the sites belonging to the largest cluster but not the shortest path are shown in blue (dark gray), and the sites belonging to other clusters than the largest one are shown in green (light gray). 
       Notice that the shortest path in the largest cluster is independent on the distribution and on the disorder strength.
       Furthermore, since for the optimal path search only nearest neighbors are considered to be connected, the optimal path crack clusters are connected via nearest and/or next-nearest neighbors.
       The optimal path crack clusters are trees by construction.
       For these pictures, the linear system size is $256$.
	}
\end{figure*}

In Fig.~\ref{fig::cluster_samples} we show snapshots of the OPC for uncorrelated lattices with different degree of disorder.
The same sequence of uniformly distributed numbers was used for all configurations.
Although the distribution of cracks and the number of dead ends are significantly affected by the degree of disorder, the shortest path in the main crack does not change. For uncorrelated lattices, the OPC shortest path is always a self-similar object with the same fractal dimension regardless the degree of disorder.

To find the mass of the shortest path in the largest cluster, the burning method introduced by Herrmann \emph{et al.}\ \cite{Herrmann84} is used. 
Since for the optimal path only the four nearest neighbors are considered to be connected and an OP consists of connected occupied sites, it is sufficient for the OPC to be connected by next-nearest neighbors. 
Therefore, for the investigation of the OPC properties, each site is considered to be directly connected to its four nearest and four next-nearest neighbors (in 3D six nearest and eight next-nearest neighbors).

The three masses introduced above are expected to asymptotically scale with the linear system size, $L$,

\begin{equation}\label{eqn::mass_exps}
	M_\text{tot} \sim L^{d_\text{tot}} \ \ , \ \ M_\text{lc} \sim L^{d_\text{lc}} \ \ , \ \ M_\text{sp} \sim L^{d_\text{sp}} \ \ ,
\end{equation}

\noindent where

\begin{equation}\label{eqn::beta_nu}
	d_\text{lc} = d - \frac{\beta}{\nu} \ \ ,
\end{equation}

\noindent is the fractal dimension of the largest cluster, $\beta$ is the critical exponent related to the order parameter, and $\nu$ is the exponent related to the correlation length. 
In addition, we measure the second moment of the OPC cluster size distribution $n_s$

\begin{equation}
	M_2 = \sum_s s^2 n_s = \frac{1}{N} \sum_k s_k^2 \ \ ,
\end{equation}

\noindent where $n_s$ is the number of clusters of size $s$ per lattice site, $N$, and $s_k$ is the size of cluster $k$. 
Percolation theory predicts that the second moment, excluding the contribution from the largest cluster, scales asymptotically as

\begin{equation}\label{eqn::second_moment_ml}
	M_2' = M_2 - \frac{\langle s_\text{max}^2 \rangle}{N} \sim L^{\gamma/\nu} \ \ ,
\end{equation}

\noindent where $s_\text{max}$ is the size of the largest OPC cluster and $\gamma$ is a critical exponent. 
For $d$ not larger than the upper critical dimension, the hyperscaling relation,

\begin{equation}\label{eqn::hyperscaling}
	d = \frac{\gamma}{\nu} + 2\frac{\beta}{\nu} \ \ ,
\end{equation}

\noindent holds. 
The scaling behavior of the OPC cluster size distribution gives access to the Fisher exponent $\tau$,

\begin{equation}
	n_s \sim s^{-\tau} \ \ .
\end{equation}

\noindent The scaling relation between $\beta$, $\gamma$, and $\tau$ reads \cite{Stauffer94}

\begin{equation}\label{eqn::scaling_b_g_t}
	\frac{\beta}{\gamma} = \frac{\tau -2}{3- \tau} \ \ .
\end{equation}

Andrade \emph{et al.}\ \cite{Andrade09} showed that the OPC depends on the value of the disorder parameter $\beta_D$.
Small values of $\beta_D$, e.\,g., $\beta_D < 1$, lead to narrow energy distributions and thus weak disorder.

\section{\label{sec::uncorrelated_lattices}Uncorrelated Lattices}

\begin{figure}
  \includegraphics[width=\columnwidth]{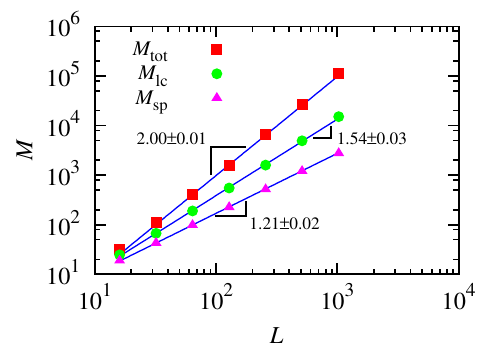}
  \caption{
    (Color online) Double logarithmic plot of the total mass, $M_\text{tot}$ ($\blacksquare$), mass of the largest cluster, $M_\text{lc}$ ($\bullet$), and mass of the shortest path, $M_\text{sp}$ ($\blacktriangle$), as a function of the linear system size, $L$, for OPC with uniformly distributed site energies. 
   This energy distribution is equivalent to the weak disorder limit. 
   The three masses scale with the system size according to three different exponents: $d_\text{tot}=2.0 \pm 0.01$, $d_\text{lc}=1.54 \pm 0.03$, and $d_\text{sp}=1.21\pm 0.02$. 
   For weak disorder, the OPC consists of a largest cluster with dangling ends and many small isolated clusters (see Fig.\,\ref{fig::sample_0_002} and (b)). 
   Results have been averaged over $10^{8}$ samples for the smallest system size and $10^{2}$ for the largest one. 
   The error bars of the individual points are smaller than the symbols.
    \label{fig::fop_random}
  }
\end{figure}

For uncorrelated OPC with uniform energy distribution, the total mass of the OPC, $M_\text{tot}$, the mass of the largest cluster of blocked sites, $M_\text{lc}$, and the mass of the shortest path, $M_\text{sp}$, as a function of the linear system size are shown in Fig.\,\ref{fig::fop_random}. 
As proposed in Eq.\,(\ref{eqn::mass_exps}), the functional relation of the masses with the system size is, within the error bars, given by power laws. 
The obtained exponents are $d_\text{tot}=2.00 \pm 0.01$, $d_\text{lc}=1.54 \pm 0.03$, and $d_\text{sp}=1.21 \pm 0.02$, consistent to the ones for $\beta_D=0.002$ \cite{Andrade09}. 
Thus, in the weak disorder regime, the three masses scale with different exponents. 
Furthermore, since the mass of all blocked sites scales linearly with the number of sites in the system, the density of cracks can be computed giving $\rho = 0.10 \pm 0.03$. 
When, instead of the site with highest energy, the one with \emph{lowest} energy is removed, the same results as for OPC in the weak disorder regime are obtained, independently of the value of $\beta_D$.

\begin{figure}
  \includegraphics[width=\columnwidth]{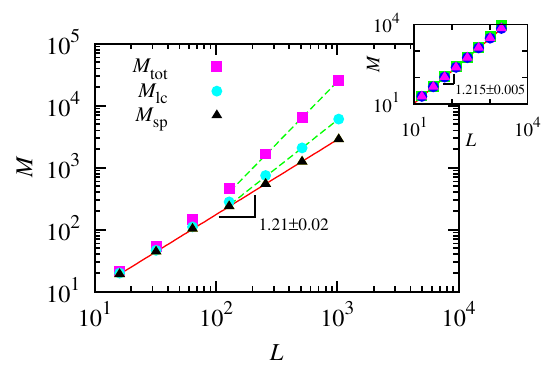}
  \caption{
    (Color online) Double logarithmic plot of the total mass, $M_\text{tot}$ ($\blacksquare$), mass of the largest cluster, $M_\text{lc}$ ($\bullet$), and mass of the shortest path, $M_\text{sp}$ ($\blacktriangle$), as a function of the linear system size, $L$, for site energies distributed according to a power-law with disorder parameter $\beta_D = 6$.
    The behavior of the total mass and of the mass of the largest cluster reveals the crossover from strong disorder to weak disorder. 
    For small system sizes -- equivalent to high disorder -- all three masses coincide, while for increasing system size -- decreasing disorder -- three different exponents emerge, as indicated by the asymptotic dashed lines which have slopes of $2.00$ (for $M_\text{tot}$) and $1.54$ (for $M_\text{lc}$). 
    The solid straight line is a guide to the eye with a slope of $d_\text{sp}=1.21\pm 0.02$.
    For this intermediate value of the disorder there are less isolated clusters than for weak disorder and the largest cluster seems to be dominated by its shortest path (see Fig.\,\ref{fig::sample_6}).
	The inset shows the same masses as the main plot, measured for a higher value of the disorder parameter: $\beta_D = 30$. For almost the entire range of system sizes the three masses are identical, indicating that these systems are in the strong disorder regime (see Fig.\,\ref{fig::sample_100}). The slope of the line gives $d_\text{sp} = 1.215 \pm 0.005$.
    Results have been averaged over $10^{8}$ samples for the smallest system size and $10^{2}$ for the largest one. 
    The error bars of the individual points are smaller than the symbols.
    \label{fig::fop_6}
  }
\end{figure}

The variation of the crossover with system size from weak to strong disorder can be observed in Fig.\,\ref{fig::fop_6}, through the behavior of the three masses as a function of the linear system size. 
Results are for the intermediate value of $\beta_D=6$. 
One can observe that for small system sizes the system is in the strong disorder regime and the three masses coincide.
The crack is then localized in a single line.
For increasing $L$ the three curves separate and tend towards their weak disorder behavior.
Andrade \emph{et al.}\ \cite{Andrade09} showed that the system size where the crossover occurs, $L_{\times}$, scales with the disorder parameter: $L_{\times}\sim \beta_D^{1/a}$, with $a\approx 0.59$.
In contrast to the behavior of $M_\text{tot}$ and $M_\text{lc}$, the mass of the shortest path in the largest cluster is independent of the system disorder and scales with the same exponent of $d_\text{sp}=1.21 \pm 0.02$ as for $\beta_D=0.002$ and for uniform energy distribution.

Three main features characterize the qualitative behavior of OPC clusters in the weak disorder limit. 
Firstly, when searching the OP, only nearest neighbors are considered to be connected and the OPC growth ends when there is no path connecting opposite sites of the system. 
Thus the OPC clusters are connected via nearest and next-nearest neighbors. 
Secondly, by construction, every site in an OPC cluster was, at some point, part of an OP, therefore there can be no extended loops which would ``trap'' sites not belonging to the OPC. 
Compact loops, i.\,e., four OPC sites at lattice positions $(x_0,y_0)$, $(x_0,y_0+1)$, $(x_0+1,y_0)$, and $(x_0+1,y_0+1)$, can, however, arise in OPC clusters. 
An example for the tree-like structure of OPC clusters is shown in Fig.\,\ref{fig::sample_0_002}. 
Finally, branching in OPC clusters is less frequent than, e.\,g., in loop-less standard or invasion percolation clusters \cite{Wilkinson83,Tzschichholz89,Cieplak96,Porto97}. 
These properties cause the shortest path, in the largest OPC cluster, to be identical to the ordinary backbone in the strong disorder limit, where the backbone becomes localized (see Fig.\,\ref{fig::sample_100}).

Next we discuss further details for the weak and strong disorder limits.

    \subsection{\label{sec::WeakDisorder}Weak Disorder}

	 \begin{figure}
	   \includegraphics[width=\columnwidth]{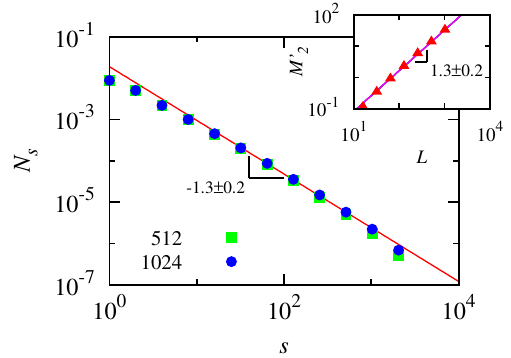}
	   \caption{
	     (Color online) Double logarithmic plot of the cluster size distribution and its second moment for energies uniformly distributed.
             The main figure shows $N_s=\sum_{r=s}^{2s-1}n_r$ ($s=1,2,4,\dots$) as a function of the cluster size, $s$, for linear system sizes $L=512$ ($\blacksquare$) and $L=1024$ ($\bullet$).
             The slope of the line gives $1-\tau = -1.3\pm 0.2$. 
             In the inset we see the second moment, $M_2'$ ($\blacktriangle$), of the cluster size distribution as a function of the system size, $L$.
             The slope of the line gives $\gamma/\nu=1.3\pm 0.2$. 
             Results have been averaged over $10^{8}$ samples for the smallest system size and $10^{2}$ for the largest one. 
             The error bars of the individual points are smaller than the symbols.
	     \label{fig::fop_tau}
	   }
	 \end{figure}

Let us analyze the properties of uncorrelated OPC clusters in the weak disorder limit. 
The critical exponents are estimated using the data for uniform site energy distribution.

In the previous section we obtained, using Eq.\,(\ref{eqn::beta_nu}), $\beta/\nu = 0.46\pm 0.03$. 
According to Eq.\,(\ref{eqn::second_moment_ml}), the ratio $\gamma/\nu$ can be determined from the scaling behavior of the second moment of the cluster size distribution, excluding the contribution of the largest cluster; the corresponding plot is shown in the inset of Fig.\,\ref{fig::fop_tau}. 
Asymptotically, the points follow a power law with exponent $\gamma/\nu = 1.3 \pm 0.2$. 
Within the error bars, the obtained results fulfill the hyperscaling relation given by Eq.\,(\ref{eqn::hyperscaling}).

From the behavior of the cluster size distribution for large cluster sizes we can extract the Fisher exponent $\tau$, as shown in Fig.\,\ref{fig::fop_tau}.
The points can be fitted by a line with slope $-1.3\pm 0.2$, such that $\tau = 2.3\pm 0.2$ (details in the caption). 
This is consistent with the obtained values for $\beta/\nu$ and $\gamma/\nu$ and the scaling relation Eq.\,(\ref{eqn::scaling_b_g_t}).

    \subsection{\label{sec::StrongDisorder}Strong Disorder}
	
For strong disorder, in contrast to weak disorder, the OPC cluster consists, mainly, in a localized crack, such that $M_\text{tot}\approx M_\text{lc}\approx M_\text{sp}$ (see Fig.\,\ref{fig::sample_100}). 
As an example, the mass scaling in OPC, with site energies distributed according to Eq.\,(\ref{eqn::energy_uncorr}) and $\beta_D = 30$, is shown in the inset of Fig.\,\ref{fig::fop_6}. 
For the smaller system sizes the three masses coincide, while for increasing system sizes one can notice a crossover of the total mass, $M_\text{tot}$. 
Since $M_\text{tot}$ grows more slowly for higher disorder (with exponent $\approx 1.2$ instead of $\approx 2.0$), the amount of OPs that have to be found to produce the OPC is smaller than in the weak disorder limit. 
Since the multiple OP searches correspond to the most time consuming part of the algorithm, the required computation effort tends to decrease with increasing disorder.
This in turn allows to determine the fractal dimension of the shortest path in the largest cluster with higher precision; we have found $d_\text{sp}=1.215\pm 0.005$, which is, within the error bars, identical to the OP fractal dimension $d_\text{op}=1.215\pm 0.005$.

\begin{figure}
	\includegraphics[width=\columnwidth]{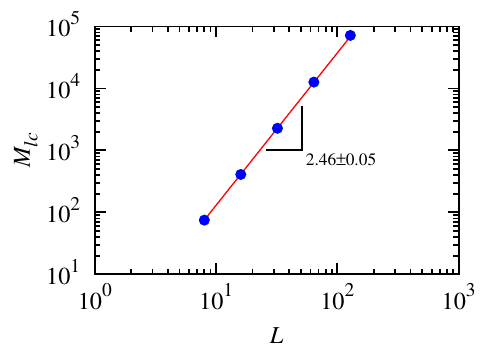}
	\caption{
	(Color online) Double logarithmic plot of the mass of the largest cluster, $M_\text{lc}$ $(\bullet)$, for uncorrelated 3D optimal path cracks, as a function of the linear system size, $L$.
        The slope of the line gives $d_\text{lc}=2.46\pm 0.05$. 
        A snapshot of one realization is shown in Fig.\,\ref{fig::fop_3d_samp}. 
        Results have been averaged over $10^{8}$ samples for the smallest system size and $50$ for the largest one. 
        The error bars of the individual points are smaller than the symbols.
	\label{fig::fop_3d}
	}
\end{figure}

\begin{figure}
	\includegraphics[width=\columnwidth]{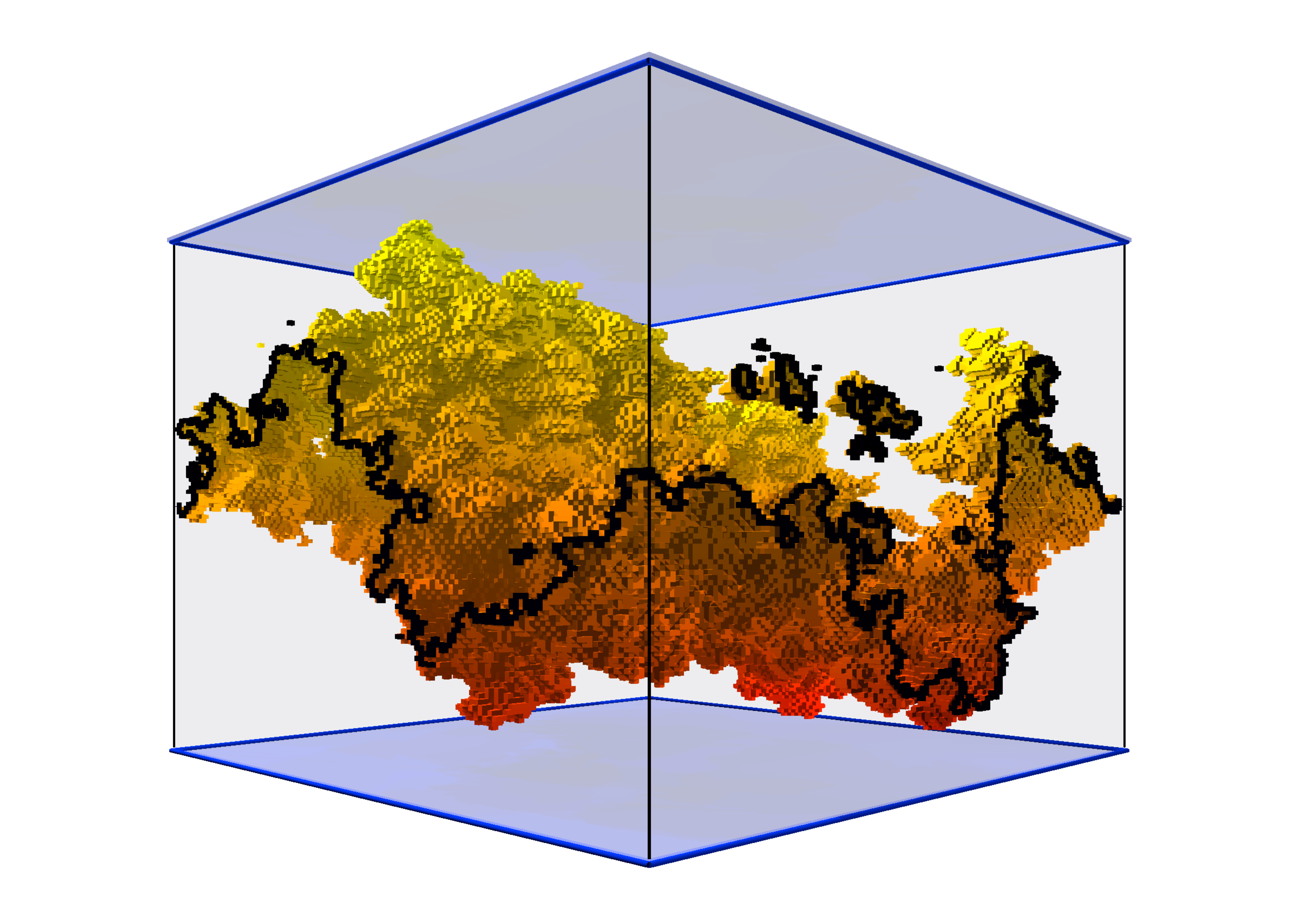}
	\caption{
	(Color online) Snapshot of a representative configuration of the optimal path crack (OPC) in three dimensions, for a system with $128^3$ lattice sites. 
        Energies have been assigned to each site in the system according to a truncated power-law with $\beta_D=100$. 
        In black are the sites where the OPC intercepts the borders of the box.
	\label{fig::fop_3d_samp}
	}
\end{figure}

We also consider the uncorrelated OPC in three dimensions, on a simple cubic lattice. 
Geometrically, it is clear that while a 2D system can be disconnected by a line, a surface is needed in 3D. 
Therefore, in analogy to the 2D case, one expects $d_\text{lc}$ to lie between two and three. 
For site energies distributed according to Eq.\,(\ref{eqn::power_dens}) with $\beta_D=100$, the mass of the largest OPC cluster as a function of the linear system size is shown in Fig.\,\ref{fig::fop_3d}. 
The obtained exponent is $d_\text{lc}=2.46\pm 0.05$. 
A snapshot of an OPC, in three dimensions, in the strong disorder regime, is shown in Fig.\,\ref{fig::fop_3d_samp}.
Alike the two-dimensional case, for three dimensions also two different regimes are observed. While for the weak disorder regime several isolated clusters are found, in the strong disorder one the cracks become localized in a single surface. However, computational limitations solely allow to accurately study the latter case. Simulations of systems in the weak disorder regime are actually limited to small system sizes and few samples.

  \section{\label{sec::correlated_lattices}Correlated Lattices}

The distributions of energies in the system have been, so far, considered spatially uncorrelated. 
In this section, we introduce a generalization of the OPC study, described above, to correlated energy landscapes.
These landscapes have been obtained from fractional Brownian motion (FBM) \cite{Mandelbrot68,Sahimi94a,Sahimi96,Peitgen88}.
We show that spatial long-range correlations lead to non-universal fractal dimensions.

\subsection{\label{sec::FBM_intro}Fractional Brownian Motion by Spectral Synthesis}

An ordinary Brownian motion \cite{Einstein05} is a stochastic process $B(t)$ with the properties

\begin{equation}
	\langle B(t) - B (t') \rangle = 0 \ \ 
\end{equation}

\noindent and

\begin{equation}
	\langle \left(B(t) - B(t')\right)^2 \rangle \propto \vert t - t' \vert \ \ ,
\end{equation}
where $t$ denotes time, and increments are mutually independent if, and only if, their time intervals do not overlap. 
Given a parameter $H$, called Hurst exponent, a fractional Brownian motion, $B_H(t)$, is a moving average of $dB(t)$, in which past increments of $B(t)$ are weighted by the kernel $(t-s)^{H-1/2}$ \cite{Mandelbrot68}. 
This leads to the properties
\begin{equation}
	\langle B_H(t) - B_H(t') \rangle = 0 \ \
\end{equation}

\noindent and

\begin{equation}
	\langle \left(B_H(t) - B_H(t')\right)^2 \rangle \propto \vert t - t' \vert^{2H} \ \ .
\end{equation}

For $H=1/2$ ordinary Brownian motion is recovered. 
In contrast to ordinary Brownian motion, in FBM, for $H\neq 1/2$, the increments are correlated and the range of this correlation is infinite. 
It can be shown that the correlation between two increments of a FBM, $B_H(t)$, is positive if $1/2<H<1$ and negative if $0<H<1/2$.

To use the properties of FBMs to obtain a correlated energy landscape, it is convenient to apply the Fourier filtering method (FFM), based on the spectral synthesis \cite{Peitgen88,Sahimi94a,Sahimi96}. 
In a nutshell, the idea is to generate random Fourier coefficients, distributed according to a given density, and to subsequently apply an inverse Fourier transform to obtain the energy landscape in the spatial domain.
It is known that, in one dimension, $1/f$ noise -- a process with spectral density $S(f)\propto 1/f^{\beta_C}$ -- is equivalent to FBM with $H=(\beta_C-1)/2$. 
Furthermore, it can also be shown that the spectral density of an FBM in $d$ dimensions can be written as
\begin{align}\label{eqn::fbm_spectrum}
	S(f_1,\dots,f_d) = \left(\sqrt{ \sum_{i=1}^d f_i^2 }\right)^{-(2H+d)} \ \ ,
\end{align}
such that $\beta_C = 2H+d$. 
Therefore, in 2D, for $2<\beta_C<3$ the increments are anti-correlated, $\beta_C = 3$ corresponds to $H=1/2$ (ordinary Brownian motion), and $3<\beta_C < 4$ leads to positively correlated increments. 
For white noise, the spectral density $S$ is constant, i.\,e., $\beta_C = 0$ or $H=-d/2$, and both the increments and the obtained lattices are uncorrelated.

In practical terms, we obtain correlated site energies in the following way \cite{Peitgen88}. 
For the spectral synthesis, Fourier coefficients corresponding to the spectral density of Eq.\,(\ref{eqn::fbm_spectrum}) are needed.
The given spectrum translates into conditions for the expectation of the absolute values of the Fourier coefficients. 
We generate these Fourier coefficients in reciprocal space through one random phase (in $[0,2\pi)$) and one amplitude distributed according to a normal distribution and with maximum amplitude
	${(\sqrt{k_1^2+k_2^2})^{-\beta_C}}$,
where $k_i$ are the frequency indices of the discrete Fourier transform.
After applying the inverse Fourier transform, we have to normalize in the range from zero to one the spatial domain distribution to represent the correlated topology. 
Separately, we generate several samples and compute the average variance $\sigma$. 
Afterwards, for each sample, we truncate its correlated distribution such that: if the distribution value is smaller (larger) than $-3\sigma$ ($3\sigma$), we assign this energy value equal to $-3\sigma$ ($3\sigma$).
This truncation affects less than $0.3\%$ of the distribution.

    \subsection{\label{sec::CorrelatedOPC}Correlated Optimal Path Cracking}

\begin{figure}
  \includegraphics[width=\columnwidth]{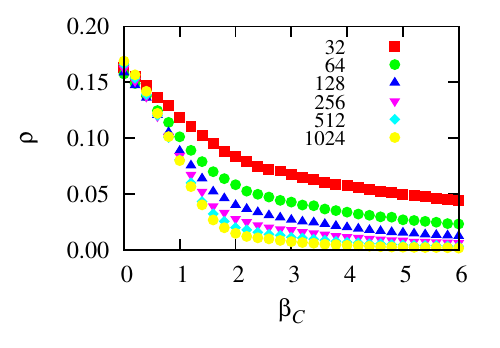}
  \caption{(Color online) Density of all removed sites for correlated OPC as a function of the correlation parameter, $\beta_C$. 
           For $\beta_C=1$, $H=-d/2$, the FBM corresponds to white noise, i.\,e., 
     		the uncorrelated case where the density is independent of the system size. 
           Data for the following system sizes are shown: $L=32$ $(\blacksquare)$, $L=64$ $(\bullet)$, $L=128$ $(\blacktriangle)$, $L=256$ $(\blacktriangledown)$, $L=512$ $(\blacklozenge)$, and $L=1024$ $(\bullet)$. 
           Results have been averaged over $3.2\times 10^3$ samples for the smallest system size and $10^2$ for the largest one. 
           The error bars of the individual points are smaller than the symbols.} 
           \label{fig::density_corr}
\end{figure}

In this section, we study the properties of OPC on correlated lattices. 
Energies distributed equivalently to the spectral density in Eq.\,(\ref{eqn::fbm_spectrum}) follow a FBM where $\beta_C$ is the spectral exponent. 
Figure \ref{fig::density_corr} shows the density of removed sites, $\rho=M_\text{tot}/N$, as a function of the correlation exponent $\beta_C$.
The number of sites that need to fail to break the global connectivity in the system decreases with the degree of correlation.
The stronger the correlations, the lower the density of removed sites.
While in the uncorrelated case ($\beta_C=0$), the density of removed sites shows no significant finite-size effects, when spatial correlations are taken into account, the larger the system size, the lower the density of removed sites.
 
\begin{figure}
  \includegraphics[width=\columnwidth]{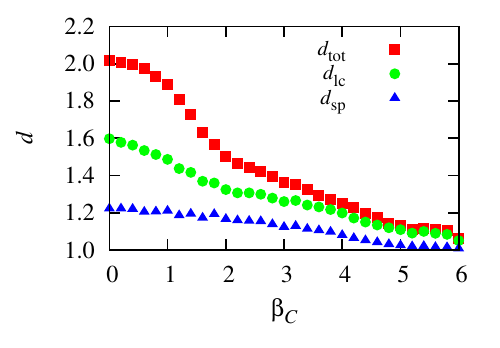}
  \caption{(Color online) The fractal dimensions $d_\text{tot}$ ($\blacksquare$), $d_\text{lc}$ ($\bullet$), and $d_\text{sp}$ ($\blacktriangle$) as a function of the correlation parameter, $\beta_C$. 
           The plot shows that the long-range spatial correlation changes the exponents of the system. 
           Results have been averaged over $3.2\times 10^3$ samples for the smallest system size and $10^2$ for the largest one. 
           The error bars of the individual points are smaller than the symbols.}
  \label{fig::df_corr}
\end{figure}

\begin{figure*}
	\centering
	\subfigure[]{
		\includegraphics[width=\columnwidth]{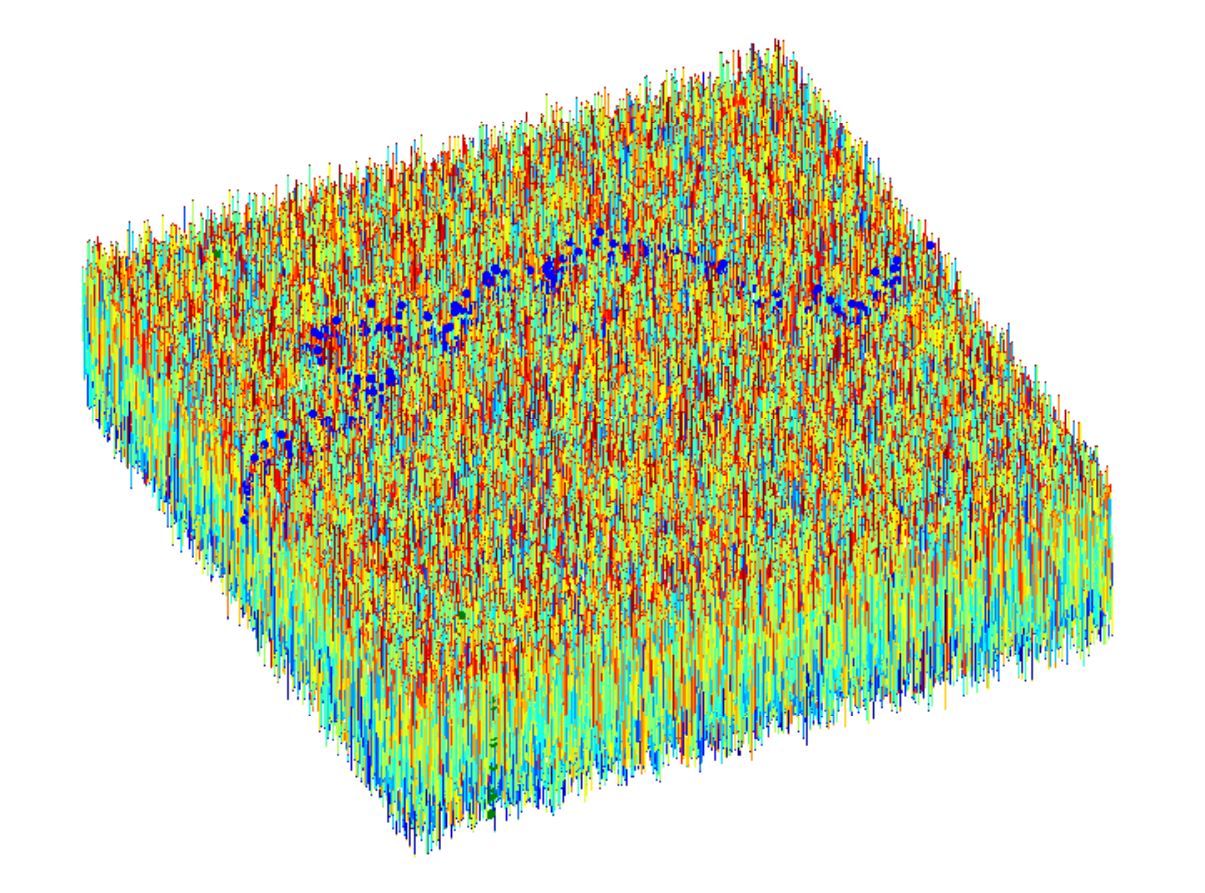}
	}
	\subfigure[]{
		\includegraphics[width=\columnwidth]{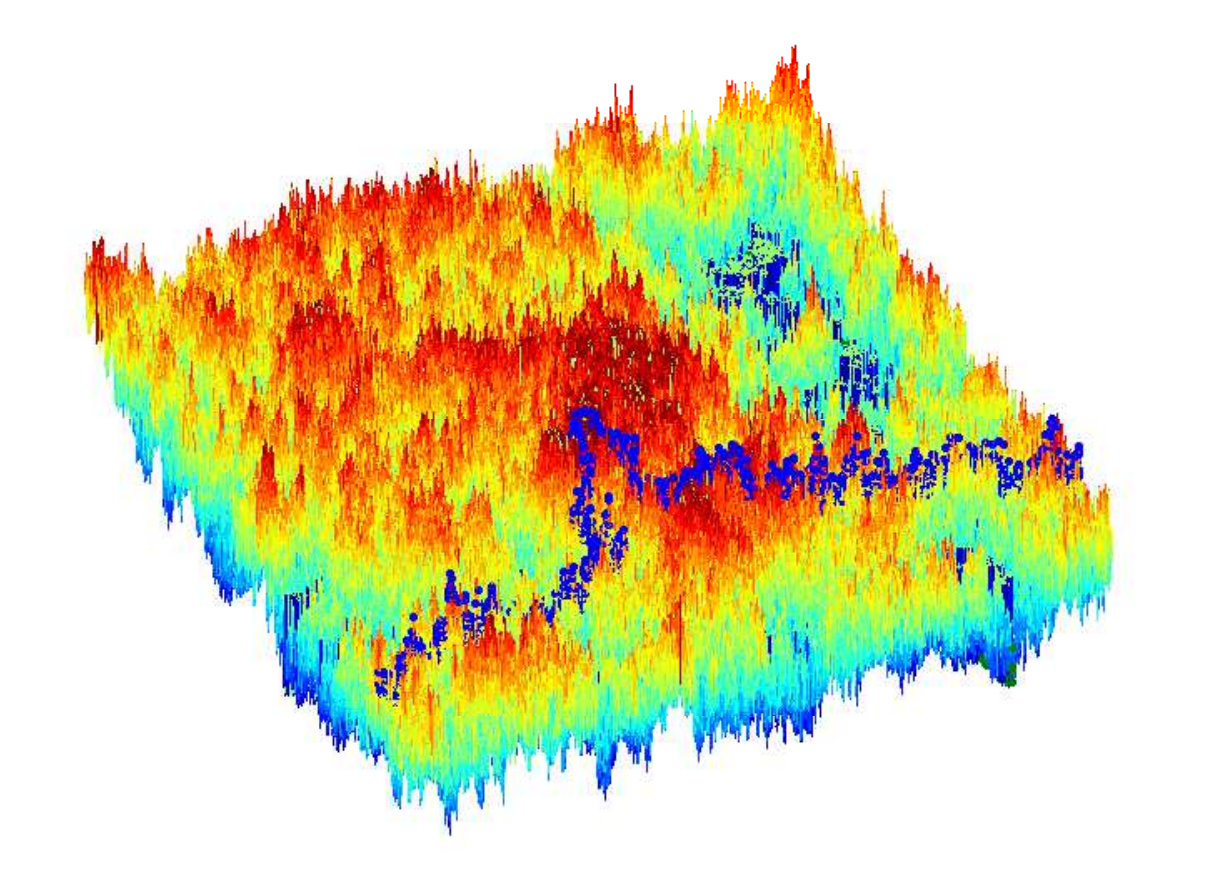}
	}
	\subfigure[]{
		\includegraphics[width=\columnwidth]{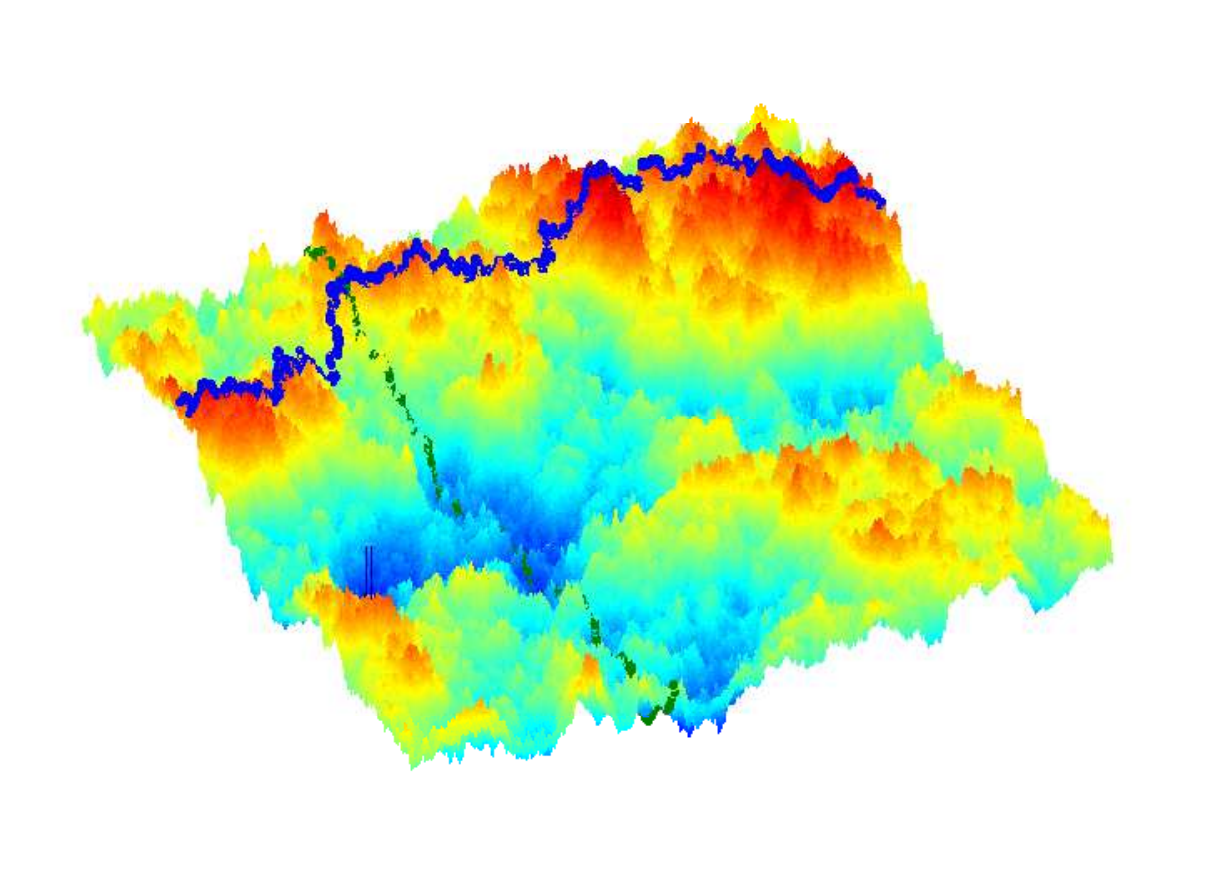}
	}
	\subfigure[]{
		\includegraphics[width=\columnwidth]{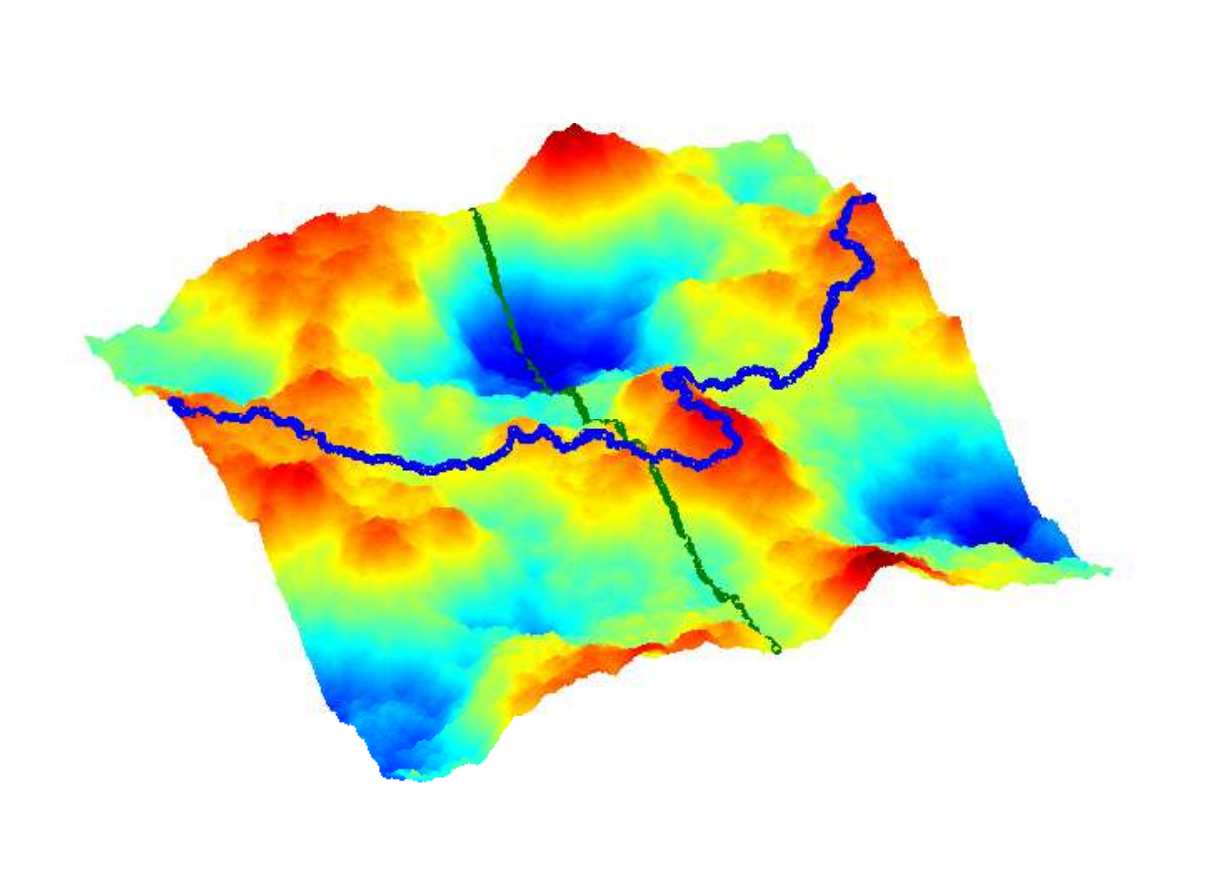}
	}
	\caption{(Color online) The shortest path in the optimal path crack for correlated and uncorrelated energy landscapes. 
                 At each lattice site, given by two coordinates $(x,y)$, the third coordinate is the energy at this point. 
                 Red regions (mountains), have energy close to unity, while blue regions (valleys) have energy close to zero. 
                 The shortest path in the optimal path crack is shown in dark blue (dark gray, left to right) and the first optimal path in dark green (medium gray, front to back). 
                 Four different correlation parameters, $\beta_C$, have been considered: a) $\beta_C=0$ $(H=-d/2)$ corresponding to the uncorrelated case -- white noise; b)  $\beta_C=2.2$ (anti-correlated); c) $\beta_C= 3$ $(H=1/2)$ recovering the ordinary Brownian motion where the increments (but not the lattice) are uncorrelated; d) $\beta_C=3.8$, corresponding to strongly correlated lattices. While the crack passes mainly through the mountain tops, the first optimal path passes mainly through the valleys -- hidden in the noisy configurations (a) and (b). For these pictures, the linear system size is 1024.
	\label{fig::corr_snap}
	}
\end{figure*}

The dependence of the exponents, $d_\text{tot}$, $d_\text{lc}$, and $d_\text{sp}$, on $\beta_C$ is shown in Fig.\,\ref{fig::df_corr}.
This plot shows that the critical exponents of correlated OPC are non-universal and change with the correlation parameter $\beta_C$.
In the absence of correlations, the values of the exponents discussed before are recovered.
For $\beta_C>0$, the exponents decrease monotonically with the correlation parameter.
No significant differences are found between correlated and anti-correlated regimes.
In the limit of strong long-range correlations, for sufficiently high values of $\beta_C$ the exponents seem to converge towards unity.

The density of all removed sites decreases monotonically with $\beta_C$ and the exponents for the different masses converge to unity, for large values of $\beta_C$.
Therefore, the OPC becomes localized in a single line.

In Fig.~\ref{fig::corr_snap} we see snapshots of the OPC shortest path and the first OP in four different energy landscapes, where the third dimension is the energy of the site.
The snapshot (a) is for the uncorrelated case.
Both the OP and the OPC have many upward and downward tilts.
Increasing the correlation, (a)--(d), the OP tends to cross the valleys and the OPC tends to go mainly through the mountains.
Even for the anti-correlated case (b), though a rough landscape is also obtained, mountains and valleys can be observed.

  \section{\label{sec::summary}Final Remarks}

In this work we have studied in detail the optimal path crack (OPC), recently introduced by Andrade \emph{et al.}\ \cite{Andrade09}, for the weak and the strong disorder limits. 
An accurate value for the fractal dimension of the shortest path in the crack was obtained. 
We also computed, for the first time, the set of critical exponents, $\beta/\nu$, $\gamma/\nu$, and $\tau$, to get more insight on the crack percolation properties, revealing an interesting set of exponents. We note that, within the error bars, the exponents $\beta/\nu$ and $\gamma/\nu$ are consistent with the ones for the parallel direction of directed percolation \cite{Henkel08}. The Fisher exponent, $\tau$, is compatible with the one for the cluster size distribution of subcritical invasion percolation clusters between two sites \cite{Araujo05} and the exponent of the distribution of areas in perturbed watersheds \cite{Fehr11}.
The OPC has also been analyzed for a three-dimensional system where the connectivity between opposite borders of the system is broken by a surface with a fractal dimension of $2.46 \pm 0.05$.

We generalized the OPC on correlated energy landscapes generated by fractional Brownian motion with different values of the correlation parameter $\beta_C$. 
For different correlations non-universal exponents are obtained for the fractal dimension of the total mass of the crack, the size of the main crack, and the length of the shortest path. 
Moreover, a monotonic decrease of the exponents with the correlation parameter is observed together with a strong tendency to localize the crack in its shortest path.
An interesting extension of this work would be to consider, for example, in the three-dimensional case, the role of correlations and to compute $\gamma/\nu$ and $\tau$.
The interplay between degree of disorder and the presence of long-range spatial correlations is still an open question. Besides, since the obtained fractal dimension of the OP and of the shortest path in the largest OPC cluster was found in several different models, it would be interesting to understand what would be the meaning of $\gamma/\nu$ and $\tau$ for these models.

\begin{acknowledgments}
We acknowledge financial support from the ETH Competence Center Coping with Crises in Complex Socio-Economic Systems (CCSS) through ETH Research Grant CH1-01-08-2.
We also acknowledge the Brazilian agencies CNPq, CAPES and FUNCAP, and the Pronex grant CNPq/FUNCAP, for financial support.
\end{acknowledgments}

\bibliography{opc}

\end{document}